\documentclass[useAMS,usenatbib]{mn2e}
\usepackage{float,multicol,epsfig}
\usepackage{amssymb}

\title[How well do we understand cosmological recombination?]
{How well do we understand cosmological recombination?}

\author[W. Y. Wong, A. Moss and D. Scott]{Wan Yan Wong\thanks{E-mail:
wanyan@phas.ubc.ca}, Adam Moss\thanks{E-mail: adammoss@phas.ubc.ca}
and Douglas Scott\thanks{E-mail: dscott@phas.ubc.ca} \\
Department of Physics and Astronomy, University of British Columbia,
 6224 Agricultural Rd., Vancouver, BC, V6T 1Z1, Canada}

\begin{document}
\date{2007}

\pagerange{\pageref{firstpage}--\pageref{lastpage}} \pubyear{2007}

\maketitle

\label{firstpage}
\begin{abstract}
The major theoretical limitation for extracting cosmological 
parameters from the CMB sky lies in the precision with which
we can calculate the cosmological recombination process.  
Uncertainty in the details of hydrogen and helium recombination
could effectively increase the errors or bias the values of the
cosmological parameters derived from the {\sl Planck} satellite, 
for example.  Here we modify the cosmological recombination code
 R{\sc ecfast} by introducing one more parameter to
reproduce the recent numerical results for the speed-up of the 
helium recombination.  Together with the existing hydrogen fudge
factor, we vary these two parameters to account for the 
remaining dominant uncertainties in cosmological recombination.
By using the C{\sc osmo}MC code with {\sl Planck} forecast
data, we find that we need to determine the parameters to better 
than ten per cent for He\,{\sc i} and one per cent for H, 
in order to obtain negligible effects on the cosmological 
parameters.  For helium recombination, if the existing studies 
have calculated the ionization fraction to the 0.1 per cent level
by properly including the relevant physical processes, then we 
already have numerical calculations which are accurate enough
for {\sl Planck}.  For hydrogen, setting the fudge factor to
speed up low redshift recombination by 14 per cent appears 
to be sufficient for {\sl Planck}.  However, more work
still needs to be done to carry out comprehensive numerical
calculations of all the relevant effects for hydrogen, as well
as to check for effects which couple hydrogen and helium 
recombinaton through the radiation field.
\end{abstract}


\section{Introduction}
{\sl Planck}~\citep{Planck:2006}, the third generation 
Cosmic Microwave Background~(CMB) satellite 
will be launched in 2008; it will measure the CMB temperature 
and polarization anisotropies $C_\ell$ at multipoles $\ell=1$
to $\simeq 2500$ at much higher precision than has been possible
before.  In order to interpret these high fidelity 
experimental data, we need to have a correspondingly high 
precision theory. 
Understanding precise details of the recombination history 
is the major limiting factor in calculating the $C_\ell$ 
to better than 1 per cent accuracy.  An assessment of the level 
of this uncertainty, in the context of the expected 
{\sl Planck} capabilities, will be the subject of this paper. 

The general physical picture of cosmological recombination 
was first given by \citet{Peebles:1968} and \citet{Zeldovich:1968}.
They adopted a simple three-level atom model for hydrogen~(H), 
with a consideration of the Ly\,$\alpha$ and lowest order 2s--1s 
two-photon rates.  Thirty years later, \citet{Seager:1999km} 
performed a detailed calculation by following all the 
resonant transitions and the lowest two-photon transition
in multi-level atoms for both hydrogen and helium in a
 blackbody radiation background.  
 \citet{Lewis:2006ym} first discussed how the uncertainties
  in recombination might bias the constraints on  
cosmological parameters coming from {\sl Planck}; 
this study was mainly motivated by the effect of 
including the semi-forbidden and high-order two-photon 
transitions~\citep{Dubrovich:2005fc}, which had been
ignored in earlier calculations.

\begin{figure}
\centering
\vspace*{6cm}
\leavevmode
\includegraphics{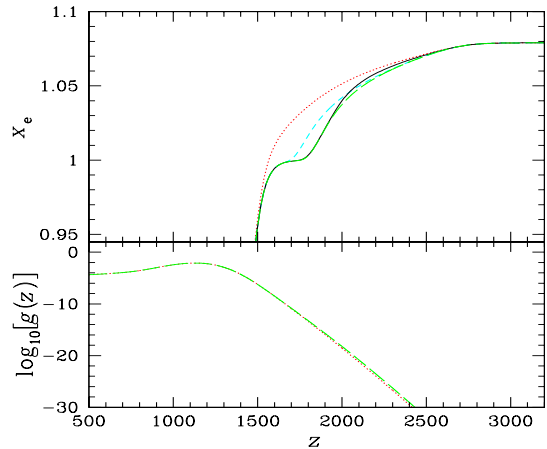}
\caption{\small Top panel: Ionization fraction $x_{\rm e}$ 
as a function of redshift $z$.
The dotted\,(red) line is calculated using the original RECFAST code.  
The solid\,(black) line is the numerical result from \citet{Switzer:2007sq},
while the dashed\,(blue) and long-dashed\,(green) lines are both 
evaluated based on the modification given by 
\citet{Kholupenko:2007qs} -- the dashed one 
has $b_{\rm He}=0.97$ (the value used in the original paper) and 
the long-dashed one has $b_{\rm He}=0.86$.
Bottom panel: The visibility function $g(z)$ versus redshift $z$.  
The two curves calculated~(dotted and long-dashed) correspond to the 
same recombination models in the upper panel.
The cosmological parameters used for these two graphs are, 
$\Omega_{\rm B} = 0.04592$, $\Omega_{\rm M} = 0.27$, 
$\Omega_{\rm CDM}= 0.22408$, $T_{\rm CMB}=2.728$\,K, 
$H_0=71$ \,kms$^{-1}$ Mpc$^{-1}$ and $f_{\rm He}=0.079$.}
\label{fig1}
\end{figure}

There have been many updates and improvements in the 
modelling of recombination since
then.  \citet{Switzer:2007sn} presented a multi-level
calculation for neutral helium~(He\,{\sc i}) recombination 
including evolution of the radiation field, which had been 
assumed to be a perfect blackbody in previous studies.
Other issues discussed recently include the continuum opacity due to 
neutral hydrogen (H)~\citep[see also][]{Kholupenko:2007qs}, 
the semi-forbidden transition $2^3$p--$1^1$s
~\citep[the possible importance of which was first proposed by][]{Dubrovich:2005fc},
the feedback from spectral distortions between $2^1$p--$1^1$s 
and $2^3$p--$1^1$s lines, and the radiative line transfer.  
In particular, continuum absorption of the $2^1$p--$1^1$s line 
photons by neutral hydrogen\,(H\,{\sc i}) causes helium recombination 
to end earlier than previously estimated~(see Fig.~\ref{fig1}).
\citet{Hirata:2007sp} also found that the high order two-photon 
rates have a  negligible effect on He\,{\sc i},
and the same conclusion was made by other groups for hydrogen 
as well~\citep{Wong:2006iv,Chluba:2007qk}, largely because the 
approximate rates adopted by~\citet{Dubrovich:2005fc} had been 
overestimated.  The biggest remaining uncertainty in He\,{\sc i}
recombination is the rate of the $2^3$p--$1^1$s transition, which causes
a variation equal to about 0.1 per cent in the ionization fraction
$x_{\rm e}$~\citep{Switzer:2007sq}.

For hydrogen, \citet{Chluba:2006bc} improved the multi-level 
calculation by considering seperate angular momentum $\ell$ 
states.  This brings about a 0.6 per cent change in $x_{\rm e}$ 
at the peak of the visibility function, and about 
1 per cent at redshifts $z < 900$.  The effect of the induced 
2s--1s two-photon rate due to the radiation background~\citep{Chluba:2005uz}
is partially compensated by the feedback of the Ly\,$\alpha$ photons
~\citep{Kholupenko:2006jm}, and the net maximum effect on $x_{\rm e}$
is only 0.55 per cent at $z \simeq 900$.  
The high-order two photon transitions bring about a 0.4 per cent
change in $x_{\rm e}$ at $z \simeq 1160$~\citep{Wong:2006iv,Chluba:2007qk}.
There are also 0.22 per cent changes in $x_{\rm e}$ at $z \simeq 1050$ when
one considers the Lyman series feedback up to $n=30$, and there is additionally
possibility of direct recombination, although this 
has only a roughly $10^{-4}$ per cent effect~\citep{Chluba:2007yp}.

The list of suggested updates on $x_{\rm e}$ is certainly not complete
yet, since some additional effects, such as the convergence of 
including higher excited states and the feedback-induced corrections 
due to the He\,{\sc i} spectral distortions,  may enhance or cancel 
other effects.  In general we still need to develop a complete 
multi-level code for hydrogen, with detailed interactions between the
atoms and the radiation field.  However, what is really important here is
establishing how these effects propagate into possible
systematic uncertainties in the estimation of cosmological 
parameters.

Since the uncertainties in cosmological recombination discussed 
in the \citet{Lewis:2006ym} paper have been reduced or updated, it is
time to revisit the topic on how the new effects or remaining 
uncertainties might affect the constraints on cosmological parameters in 
future experiments.
The recent full version of the He\,{\sc i} recombination 
calculation~\citep{Switzer:2007sn,Switzer:2007sq,Hirata:2007sp} 
takes too long to run to be included within the current
Boltzmann codes for $C_\ell$.  So instead, in this paper, we try to 
reproduce the updated ionization history by modifying 
R{\sc ecfast}~\citep{Seager:1999bc} using a simple parametrization
based on the fitting formulae provided by \citet{Kholupenko:2007qs}.
We then use the C{\sc osmo}MC~\citep{Lewis:2002ah} code to 
investigate how much this impacts the constraints on cosmological 
parameters for an experiment like {\sl Planck}.

%
\begin{figure}
\centering
\vspace*{6cm}
\leavevmode
\includegraphics{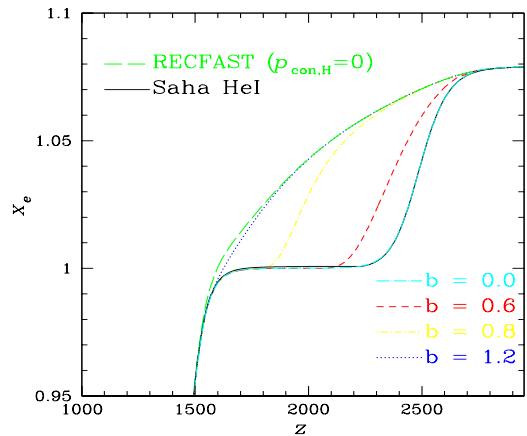}
\caption{\small Ionization fraction $x_{\rm e}$ as a function of 
redshift $z$ calculated based on the modified He\,{\sc i} recombination 
discussed here with different values of the helium fitting 
parameter $b_{\rm He}$. 
The curve with $b_{\rm He}=0$~(dot-dashed, blue) overlaps  
the line using Saha equilibrium recombination~(solid, black).
 The cosmological parameters used in this graph are
the same as for Fig.~\ref{fig1}.}
\label{fig3}
\end{figure}
%
\begin{figure}
\centering
\vspace*{6cm}
\leavevmode
\includegraphics{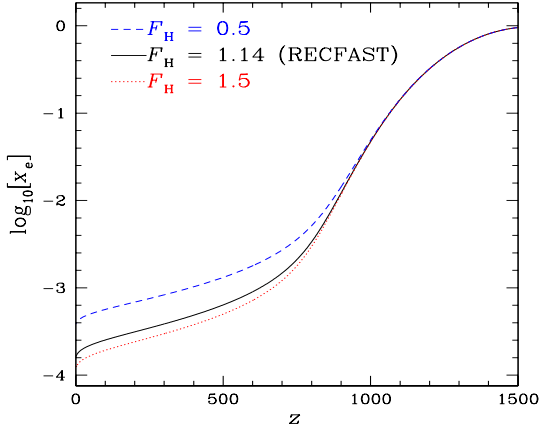}
\caption{\small The ionization fraction $x_{\rm e}$ as a function of 
redshift $z$ calculated with different values of the hydrogen fudge factor
$F_{\rm H}$.  The cosmological parameters used in this graph are
the same as in Fig.~\ref{fig1}.}
\label{fig2}
\end{figure}
%
\section{Recombination model}
In this paper, we modify R{\sc ecfast} based on the fitting 
formulae given by \citet{Kholupenko:2007qs} for including 
the effect of the continuum opacity of neutral hydrogen 
for He\,{\sc i} recombination.  The basis set of rate equations
 of the ionization fraction of H and He\,{\sc i} used in 
R{\sc ecfast} are: 
{\setlength\arraycolsep{1pt}
\begin{eqnarray}
&& H(z)(1+z) {dx_{\rm p}\over dz}  = \nonumber \\
&& \quad \quad \quad \quad  \Big(x_{\rm e}x_{\rm p} n_{\rm H} 
 \alpha_{\rm H} - \beta_{\rm H} (1-x_{\rm p}) 
   {\rm e}^{-h\nu_{\rm H2s}/kT_{\rm M}}\Big) C_{\rm H}, \\
\label{eqHeI_xe}
&& H(z)(1+z) {dx_{\rm He II}\over dz} =  \nonumber \\
 &&   \left(x_{\rm He II}x_{\rm e} n_{\rm H} \alpha_{\rm HeI}
   - \beta_{\rm HeI} (f_{\rm He}-x_{\rm He II})
   {\rm e}^{-h\nu_{{\rm HeI}, 2^1{\rm s}}/kT_{\rm M}}\right) C_{\rm HeI} \nonumber \\
 && + \left(x_{\rm He II}x_{\rm e} n_{\rm H} \alpha^{\rm t}_{\rm HeI}
   - \frac{g_{{{\rm HeI}, 2^3{\rm s}}}}{g_{{\rm HeI}, 1^1{\rm s}}} 
   \beta^{\rm t}_{\rm HeI} (f_{\rm He}-x_{\rm He II})
   {\rm e}^{-h\nu_{{\rm HeI},2^3{\rm s}}/kT_{\rm M}}\right)  \nonumber \\
&& \quad \times C_{\rm HeI}^{\rm t}  \ , 
\end{eqnarray}}
\\
where
{\setlength\arraycolsep{1pt}
\begin{eqnarray}
&& C_{\rm H} = {1 + K_{\rm H} \Lambda_{\rm H} n_{\rm H}(1-x_{\rm p})
    \over 1+K_{\rm H} (\Lambda_{\rm H} + \beta_{\rm H})
     n_{\rm H} (1-x_{\rm p}) }, \\
&& C_{\rm HeI} = {1 + K_{\rm HeI} \Lambda_{\rm He} n_{\rm H}
  (f_{\rm He}-x_{\rm He II}){\rm e}^{h\nu_{\rm ps}/kT_{\rm M}}
  \over 1+K_{\rm HeI}
  (\Lambda_{\rm He} + \beta_{\rm HeI}) n_{\rm H} (f_{\rm He}-x_{\rm He II})
  {\rm e}^{h\nu_{\rm ps}/kT_{\rm M}} },  \\
&& C_{\rm HeI}^{\rm t} = {1  \over 1+K^{\rm t}_{\rm HeI}
  \beta^{\rm t}_{\rm HeI} n_{\rm H} (f_{\rm He}-x_{\rm He II})
  {\rm e}^{h\nu^{\rm t}_{\rm ps}/kT_{\rm M}}}.
\end{eqnarray}}
\\
Note that $x_{\rm e}$ is defined as the ratio of free electons
per H atom and so $x_{\rm e} > 1$ during He recombination. 
We follow the exact notation used in~\citet{Seager:1999bc} and 
we do not repeat the definitions of all symbols, except those that
did not appear in that paper.  The last term in 
equation~(\ref{eqHeI_xe}) is added to the original  $dx_{\rm He II}/dz$
rate for the recombination of He\,{\sc i} through the triplets by 
including the semi-forbidden transition from the $2^3$p state to
the $1^1$s ground state.  This additional term can be easily 
derived by considering an extra path for 
electrons to cascade down in He\,{\sc i} by going from the continuum
through $2^3$p to ground state, and assuming that the rate of change of 
the population of the $2^3$p state is negligibly small. 
 The superscript `t' stands for triplets, so that, for example,
$\alpha^{\rm t}_{\rm HeI}$ is the Case B He\,{\sc i} recombination
for triplets.  Based on the data given by~\citet{Hummer:1998}, 
$\alpha^{\rm t}_{\rm HeI}$ is fitted with the same functional 
form used for the $\alpha_{\rm HeI}$ 
singlets~\citep[see equation~(4), in][]{Seager:1999bc},
with different values for the  parameters: 
 $p=0.761$; $q=10^{-16.306}$; $T_1=10^{5.114}$\,K; and $T_2=3$\,K. 
This fit is accurate to better than 1 per cent for temperatures between 
$10^{2.8}$ and $10^{4}$\,K. Here $\beta^{\rm t}_{\rm HeI}$ is the 
photoionization coefficient for the triplets and is calculated
from $\alpha^{\rm t}_{\rm HeI}$ by
{\setlength\arraycolsep{1pt}
\begin{eqnarray}
\beta^{\rm t}_{\rm HeI} = \alpha^{\rm t}_{\rm HeI}
\left(\frac{2 \pi m_{\rm e} k_{\rm B} T_{\rm M}}{h^2}\right)^{3/2}
\frac{2 g_{\rm He^+}}{g_{\rm HeI, 2^3s}} e^{-h \nu_{\rm 2^3s,c}/kT_{\rm M}},
\end{eqnarray}}
\\
where $g_{\rm He^+}$ and $g_{\rm HeI, 2^3s}$ are the degeneracies of
He$^+$ and of the He\,{\sc i} atom with electron in the $2^3$s state,
and $h \nu_{\rm 2^3s,c}$ is the ionization energy of the $2^3$s state.  

The correction factor $C_{\rm HeI}$ accounts for the 
slow recombination due to the bottleneck of the He\,{\sc i}
$2^1$p--$1^1$s transition among singlets.  We can also derive
the corresponding correction factor $C_{\rm HeI}^{\rm t}$
for the triplets.
The $K_{\rm H}$, $K_{\rm HeI}$ and $K^{\rm t}_{\rm HeI}$ quantities
are the cosmological redshifting of the H Ly\,$\alpha$, 
He\,{\sc i} $2^1$p--$1^1$s  and He\,{\sc i} $2^3$p--$1^1$s 
transition line photons, respectively.  The factor $K$  
used in R{\sc ecfast} is a good approximation
 when the line is optically thick ($\tau \gg 1$) and
the Sobolev escape probability $p_{\rm S}$ is roughly equal to $1/\tau$.
In general, we can relate $K$ and $p_{\rm S}$ through the following 
equations~(taking He\,{\sc i} as an example):
{\setlength\arraycolsep{1pt}
\begin{eqnarray}
&& K_{\rm HeI} = \frac{g_{{\rm HeI}, 1^1{\rm s}}}{g_{{\rm HeI}, 2^1{\rm p}}}
\frac{1}{ n_{{\rm HeI}, 1^1{\rm s}}  
A^{\rm HeI}_{ 2^1{\rm p}-1^1{\rm s}} p_{\rm S}} \quad {\rm and}\\
&& K^{\rm t}_{\rm HeI} = \frac{g_{{\rm HeI}, 1^1{\rm s}}}{g_{{\rm HeI}, 2^3{\rm p}}}
\frac{1}{ n_{{\rm HeI}, 1^1{\rm s}}  
A^{\rm HeI}_{ 2^3{\rm p}-1^1{\rm s}} p_{\rm S}} \ , 
\label{eqKHeI}
\end{eqnarray}}
\\
where 
$A_{{\rm HeI}, 2^1{\rm p}-1^1{\rm s}}$ and $A_{{\rm HeI}, 2^3{\rm p}-1^1{\rm s}}$ 
are the Einstein $A$ coefficients of the He\,{\sc I} $2^1$p--$1^1$s 
and  He\,{\sc I} $2^3$p--$1^1$s transitions, respectively.  Note 
that $A_{{\rm HeI}, 2^3{\rm p} -1^1{\rm s}}= 
g_{{\rm HeI}, 2^3{\rm P}_1}/g_{{\rm HeI}, 2^3{\rm p}}
 \times A_{{\rm HeI}, 2^3{\rm P}_1-1^1{\rm s}} 
= 1/3 \times 177.58\,$s$^{-1}$~\citep{Lach:2001}.   
For He\,{\sc i} $ 2^1$p--$1^1$s, 
we replace $p_{\rm S}$ by the new escape probability $p_{\rm esc}$, 
to include the effect of the continuum opacity due to H, 
based on the approximate formula suggested by~\citet{Kholupenko:2007qs}. 
Explicitly this is
{\setlength\arraycolsep{1pt}
\begin{eqnarray}
&& p_{\rm esc} = p_{\rm S} + p_{\rm con, H} \, , \\
{\rm where} &&  \nonumber \\
&& p_{\rm S} = \frac{1 - e^{-\tau}}{\tau} \quad {\rm and} \\
\label{pconHe}
&& p_{\rm con, H} = \frac{1}{1 + a_{\rm He} \gamma^{b_{\rm He}}}, \\
{\rm with \ } && \nonumber \\
&& \gamma = \frac{\frac{g_{{\rm HeI}, 1^1{\rm s}}}{g_{{\rm HeI}, 2^1{\rm p}}}
A^{\rm HeI}_{2^1{\rm p}-1^1{\rm s}} (f_{\rm He} - x_{\rm HeII})c^2}
{8 \pi^{3/2} \sigma_{{\rm H},1{\rm s}}(\nu_{\rm HeI,2^1 p}) 
\nu_{\rm HeI,2^1{\rm p}}^2 \Delta \nu_{\rm D,2^1p} 
(1 - x_{\rm p})}\, , \nonumber
\end{eqnarray}}
\\
where $\sigma_{{\rm H},1s}(\nu_{\rm HeI,2^1p})$ is the H ionization 
cross-section at frequency $\nu_{\rm HeI,2^1p}$ and 
$\Delta \nu_{\rm D,2^1p} = \nu_{\rm HeI,2^1p} \sqrt{2 k_{\rm B} 
T_{\rm M}/m_{\rm He} c^2}$ 
is the thermal width of the He\,{\sc i} $2^1$p--$1^1$s line.
The $\gamma$ factor in $p_{\rm con, H}$ is approximately the 
ratio of the He\,{\sc i} $2^1$p--$1^1$s transition rate to the 
H photoionization rate.  When $\gamma \gg 1$, the effect 
of the continuum opacity due to neutral hydrogen on 
the He\,{\sc i} recombination is negligible.
Here $a_{\rm He}$ and $b_{\rm He}$ are fitting parameters, which are equal to
0.36 and 0.97, based on the results from~\citet{Kholupenko:2007qs}.

We now try to reproduce these results with our modified R{\sc ecfast}.
Fig.~\ref{fig1}\,(upper panel) shows the numerical result 
of the ionization fraction $x_{\rm e}$ from different 
He\,{\sc i} recombination calculations.  The results from 
\citet{Kholupenko:2007qs} and \citet{Switzer:2007sq}
both demonstrate a significant speed up of He\,{\sc i} recombination
compared with the original R{\sc ecfast}.  We do not expect these 
two curves to match each other, since \citet{Kholupenko:2007qs}
just included the effect of the continuum opacity 
due to hydrogen, which is only one of the main improvements stated 
in \citet{Switzer:2007sq}.  
Nevertheless, we can regard the \citet{Kholupenko:2007qs} study 
as giving a simple fitting model in a three-level atom to 
account for the speed-up of the He\,{\sc i} recombination.
Fig.~\ref{fig3} shows how the ionization history changes 
with different values of the fitting parameter $b_{\rm He}$
(with $a_{\rm He}$ fixed to be 0.36).
  When $b_{\rm He}$ is larger than 1.2, the effect of the 
neutral H is tiny and the fit returns to the situation
with no continuum opacity.  However, if $b_{\rm He}$ 
is smaller than 1, the effect of the continuum opacity 
becomes more significant.  Note that when $b_{\rm He}=0$, 
both the escape probability $p_{\rm esc}$ and the
correction factor $C_{\rm HeI}$ are close to unity.
This means that almost all the emitted photons can escape
 to infinity and so the ionization history returns to Saha 
equilibrium for He\,{\sc i} recombination.  

This simple fitting formula can reproduce quite well 
the detailed numerical result for the 
ionization history at the later stages of He\,{\sc i} 
recombination.  From Fig.~\ref{fig1}, we can see 
that our model with $b_{\rm He}=0.86$ matches with the
 numerical result at $z \lesssim 2000$~\citep{Switzer:2007sq}. 
Although our fitting model does not agree so well with 
the numerical results for the earlier stages of He\,{\sc i} 
recombination, the effect on the $C_\ell$ is 
neligible.  This is because the visibility function 
$g(z) \equiv e^{- \tau} d \tau/dz$,
is very low at $z > 2000$ (at least 16 orders of magnitude 
smaller than the maximum value of $g(z)$), as shown in 
the lower panel of
Fig.~\ref{fig1}.  Our fitting approach also appears 
to work well for other cosmological 
models~(Switzer \& Hirata, private communication).  
%
\begin{figure}
\centering
\vspace*{7cm}
\leavevmode
\includegraphics{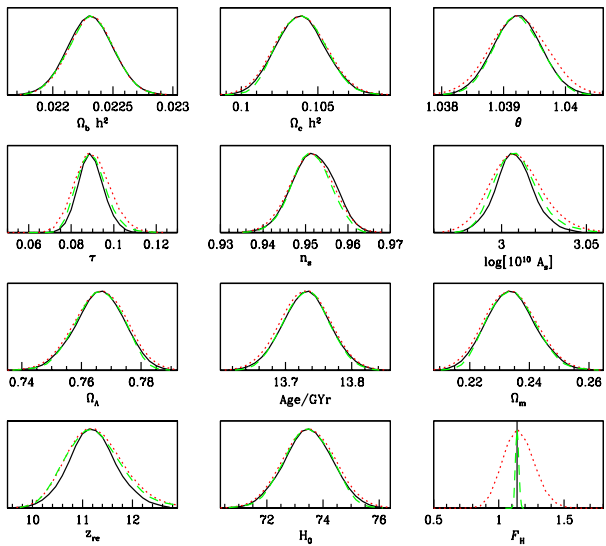}
\caption{Marginalized posterior distributions for forecast 
{\sl Planck} data varying the hydrogen recombination only.
All the curves are generated using the original R{\sc ecfast} code.
The solid\,(black) curve uses fixed $F_{\rm H}$, while the
dotted\,(red) and dashed\,(green) allow for varying $F_{\rm H}$ 
with Gaussian distributions centred at 1.14, with 
$\sigma = 0.1$ and 0.01, respectively.  Note that using a flat
prior (between 0 and 1.5) for $F_{\rm H}$ gives the same spectra 
as the case with $\sigma = 0.1$ (the red dotted line).}
\label{fig4}
\end{figure}
%
\begin{figure}
\centering
\vspace*{9cm}
\leavevmode
\includegraphics{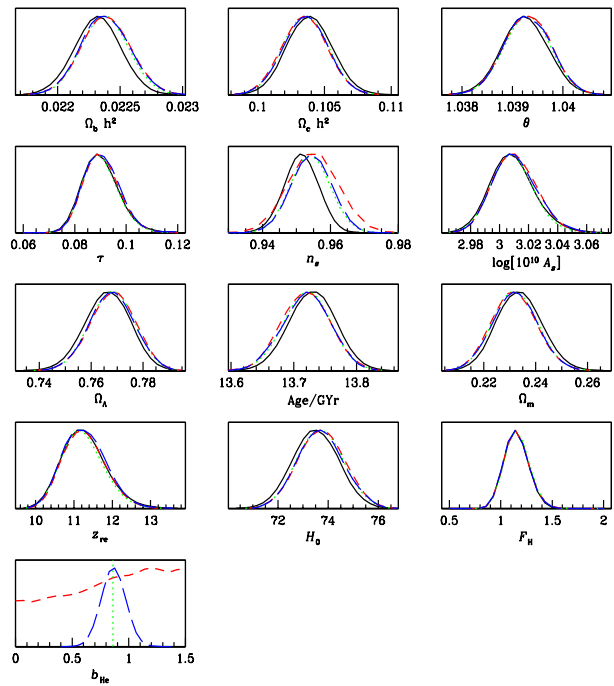}
\caption{Marginalized posterior distributions for forecast 
{\sl Planck} data with hydrogen and helium phonomenological parameters
both allowed to vary.
The solid\,(black) curve shows the constraints using the 
original R{\sc ecfast} code and allowing $F_{\rm H}$ to be  
a free parameter.  The other curves also allow for the 
variation of $F_{\rm H}$ and use the fitting function 
for He\,{\sc i} recombination described in Section 2:
the dotted\,(green) line sets $b_{\rm He}$ equal to 0.86; the dashed\,(red)
 one is with a flat prior for $b_{\rm He}$ from 0 to 1.5; and the 
long-dashed\,(blue) one is with a narrow prior for $b_{\rm He}$, consisting of 
a Gaussian centred at 0.86 and with $\sigma=0.1$.}
\label{fig5}
\end{figure}

%
In this paper, we employ the fudge factor $F_{\rm H}$ 
for H~(which is the extra factor multiplying
$\alpha_{\rm H}$) and the He\,{\sc i} parameter 
$b_{\rm He}$ in our model to represent the remaining 
uncertainties in recombination.  For He\,{\sc i}, the 
factors $a_{\rm He}$ and $b_{\rm He}$ in 
equation~(\ref{pconHe}) are quite degenerate.
We choose to fix $a_{\rm He}$ and use $b_{\rm He}$ as 
the free parameter in this paper; this is because
it measures the power dependence of the ratio of 
the relevant rates $\gamma$ in the escape probability 
due to the continuum opacity $p_{\rm con, H}$.
For hydrogen recombination, all the individual updates
suggested recently give an overall change less 
than 0.5 per cent in $x_{\rm e}$ around 
the peak of the visibility function.  Only the effect of 
considering the separate $\ell$-states causes more than
a 1 per cent change, and only for the final stages 
of hydrogen recombination ($z \lesssim 900$).
Therefore, we think it is sufficient
to represent this uncertainty with the usual fudge factor
$F_{\rm H}$, which basically controls the speed of the 
end of hydrogen recombination (see Fig.~\ref{fig2}).
The best-fit to the current recombination calculation 
has $F\simeq 1.14$.
%
\section{Forecast data}
We use the C{\sc osmo}MC~\citep{Lewis:2002ah} code to 
perform a Markov Chain Monte Carlo (MCMC) 
calculation for sampling the posterior distribution with 
given forecast data.  The simulated {\sl Planck} data and 
 likelihood function are generated based on the 
settings suggested in \citet{Lewis:2006ym}. 
We use full polarization information for {\sl Planck} by considering 
the temperature $T$ and $E$-type polarization
anisotropies for $\ell \leq 2400$, and assume that they are 
statistically isotropic and Gaussian.  The noise is 
also isotropic and is based on a simplified model with
$N^{TT}_{\ell} = N^{EE}_{\ell}/4 = 2 \times 10^{-4} \mu$K$^2$, 
having a Gaussian beam of 7 arcminutes
~\citep[Full Width Half Maximum,][]{Planck:2006}.
For our fiducial model, we adopt the best values of the six 
cosmological parameters in a $\Lambda$CDM model from the {\sl WMAP} 
three-year result~\citep{Spergel:2006hy}.
The six parameters are the baryon density 
$\Omega_{\rm b}h^2= 0.0223$, the cold dark matter density 
$\Omega_{\rm C}h^2=0.104$, the present Hubble parameter 
 $H_0= 73$\,km\,s$^{-1}$Mpc$^{-1}$, the constant scalar 
adiabatic spectral index $n_{\rm s}=0.951$, the scalar amplitude 
(at $k=0.05$\,Mpc$^{-1}$) 
$10^{10}A_{\rm s} = 3.02$ and the optical depth due to 
reionization (based on a sharp transition) $\tau = 0.09$. 
For recombination, we calculate the ionization history 
using the original R{\sc ecfast} with the fudge 
factor for hydrogen recombination $F_{\rm H}$ set to 1.14
and the helium abundance equal to 0.24.

In this study, we only vary the basic six standard cosmological
parameters stated above, together with the hydrogen fudge 
factor $F_{\rm H}$ and also the helium $b_{\rm He}$ factor 
for the recombination process.  Of course degeneracies 
will in general be worse if one allows for a wider
set of parameters.
%
\section{Results}
\begin{figure}
\centering
\vspace*{8cm}
\leavevmode
\includegraphics{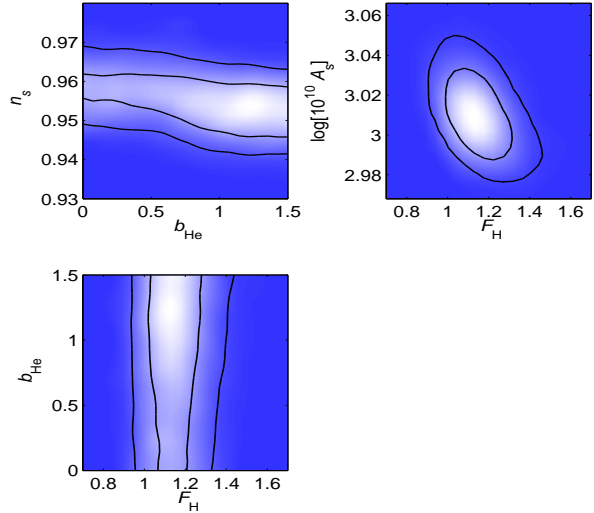}
\caption{Projected 2D likelihood for the four parameters
$n_{\rm s}$, $A_{\rm s}$, $F_{\rm H}$ and $b_{\rm He}$.
Shading corresponds to the marginalized probabilities 
with contours at 68 per cent and 95 per cent confidence.}
\label{fig6}
\end{figure}

Fig.~\ref{fig4} shows the parameter constraints from our forecast
{\sl Planck} likelihood function using the original R{\sc ecfast} 
code with varying $F_{\rm H}$ and adopting different priors.
For the {\sl Planck} forecast data, $F_{\rm H}$ can be well
constrained away from zero~\citep[the same result as in][]{Lewis:2006ym} 
and is bounded by a nearly Gaussian distribution
with $\sigma$ approximately equal to 0.1.
When we only vary $F_{\rm H}$ with different priors 
(compared with fixing it to 1.14), it basically does not change 
the size of the error bars on the cosmological parameters, 
except for the scalar adiabatic amplitude $10^{10}A_{\rm s}$.  
From Fig.~\ref{fig2}, 
we can see that the factor $F_{\rm H}$
controls the speed of the final stages of H recombination,
when most of the atoms and electrons 
have already recombined.  Changing $F_{\rm H}$ affects the 
optical depth $\tau$ from Thomson scattering, which determines 
the overall normalization amplitude of $C_\ell$ ($\propto e^{-2 \tau}$) 
at angular scales below that subtended by the size of the horizon at last
scattering~($\ell \gtrsim 100$).  This is the reason 
why varying $F_{\rm H}$ affects the uncertainty in $A_{\rm s}$, 
since $A_{\rm s}$ also controls the overall amplitude of $C_\ell$
(see the upper right panel in Fig.~\ref{fig6} for the marginalized 
distribution for $F_{\rm H}$ and $A_{\rm s}$).
The modified recombination model also changes the peak value
(but not really the width) of the adiabatic spectral index 
$n_{\rm s}$ distribution, as one can see by comparing the dotted and 
dashed curves in Fig.~\ref{fig4}.
  
Based on all the suggested effects on H recombination, the 
uncertainty in $x_{\rm e}$ is at the level of a few per cent
 at $z \lesssim 900$, which corresponds to roughly a 1
per cent change in $F_{\rm H}$.  In Fig.~\ref{fig4}, 
we have also tried to take this uncertainty into 
account by considering a prior on $F_{\rm H}$ with $\sigma=0.01$ 
(the long-dashed curves).  We find that the result is almost 
the same as for the case using $\sigma=0.1$ for the $F_{\rm H}$ prior.
On the other hand, the error bar (measured using the 68 per cent 
confidence level, say) of $A_{\rm s}$ is increased by 40 
and 16 per cent with $\sigma=0.1$ and 0.01, respectively.

Fig.~\ref{fig5} shows the comparison of the constraints in
the original and modified versions of R{\sc ecfast}, with
both H and He parameterized.  
By comparing the solid and dotted curves in Fig.~\ref{fig5}, 
we can see that only the peaks of the spectra of the cosmological 
parameters are changed, but not the width of the distributions,
when switching between the original and modified R{\sc ecfast} codes.
Allowing $b_{\rm He}$ to float in the 
modified recombination model only leads to an increase in the error bar 
for spectral index $n_{\rm s}$ among all the parameters, 
including $F_{\rm H}$.  For the dashed curves, we used a very 
conservative prior for $b_{\rm He}$, namely a flat spectrum 
from 0 to 1.5~(i.e. from Saha recombination to the old 
R{\sc ecfast} behaviour).  We can see that the value of $b_{\rm He}$ 
is poorly constrained, because the CMB is only weakly sensitive 
to the details of He\,{\sc i} recombination.  
Nevertheless, this variation allows for faster He\,{\sc i}
recombination than in the original R{\sc ecfast} code and this 
skews the distribution of $n_{\rm s}$ towards higher 
values~(see also the upper left panel in Fig.~\ref{fig6}).  
This is because a faster He\,{\sc i} recombination leads to 
fewer free electrons before H recombination and this increases 
the diffusion length of the photons and baryons.  This 
in turn decreases the damping scale of the acoustic oscillations
at high $\ell$, which therefore gives a higher value of $n_{\rm s}$.
In addition, this variation in $b_{\rm He}$ 
increases the uncertainty (at the 68 per cent confidence level) 
of $n_{\rm s}$ by 11 per cent.

Based on the comprehensive study of \citet{Switzer:2007sq},
the dominant remaining uncertainty in He\,{\sc i} recombination
is the $2^3$p--$1^1$s transition rate, which causes about 
a 0.1 per cent variation in $x_{\rm e}$ at $z \simeq 1900$.
For our fitting procedure this corresponds to about a 1 
per cent change in $b_{\rm He}$.  We try to take this 
uncertainty into account in our calculation by adopting 
a prior on $b_{\rm He}$ which is peaked at 0.86 with width~(sigma) 
liberally set to 0.1.  From Fig.~\ref{fig5}, one can see that the 
error bar on $n_{\rm s}$ is then reduced to almost the 
same size as found when fixing $b_{\rm He}$ equal to 0.86 
(the dotted and long-dashed curves).  This means that, for
the sensitivity expected from {\sl Planck},
it is sufficient if we can determine $b_{\rm He}$ 
to better than 10 per cent accuracy.

As well as the individual marginalized uncertainties,
we can also look at whether there are degeneracies among the 
parameters.  From Fig.~\ref{fig6}, we see that $F_{\rm H}$ and 
$b_{\rm He}$ are quite independent.
This is because the two parameters govern recombination 
at very different times.  As discussed before, $b_{\rm He}$
 controls the speed of He\,{\sc i} recombination, which affects 
the high-$z$ tail of the visiblity function, while $F_{\rm H}$ 
controls the low-$z$ part.
%
%
\section{Discussion and Conclusions}
In this paper, we modify R{\sc ecfast} by introducing
one more parameter $b_{\rm He}$ (besides the hydrogen fudge 
factor $F_{\rm H}$) to mimic the recent numerical results for 
the speed-up of He\,{\sc i} recombination.  
By using the C{\sc osmo}MC code with forecast {\sl Planck}
data, we examine the variation of these two factors to 
account for the remaining dominant uncertainties 
in the cosmological recombination calculation.  For He\,{\sc i}, 
the main uncertainty comes from the $2^3$p--$1^1$s 
rate~\citep{Switzer:2007sq}, which corresponds to 
about a 1 per cent change in $b_{\rm He}$.  We find that
this level of variation has a negligible effect on the 
determination of the cosmological parameters.  Therefore, 
based on this simple model, if the existing studies have 
properly considered all the relevant physical radiative processes 
in order to provide $x_{\rm e}$ to 0.1 per cent accuracy during 
He\,{\sc i} recombination, then we already have numerical
 calculations which are accurate enough for {\sl Planck}.

For H, since there is still no comprehensive 
model which considers all the interactions between
the atomic transitions and the radiation background,
we consider the size of the updates as 
an indication of the existing level of uncertainty.
We represent this uncertainty by varying the 
fudge factor $F_{\rm H}$, because the largest update on 
$x_{\rm e}$ occurs at $z \lesssim 900$, and comes from  
a consideration of the separate angular momentum
states~\citep{Chluba:2006bc}.  We find that
$F_{\rm H}$ needs to be determined to better than 
1 per cent accuracy in order to have negligible 
effect on the determination of cosmological parameters
with {\sl Planck}. 

 Hydrogen recombination is of course
important for the formation of the CMB anisotropies 
$C_{\ell}$, since it determines the detailed profile 
around the peak of the visibility function $g(z)$.
A comprehensive numerical calculation of the recombination of H
(similar to He\,{\sc i}) to include at least all the recent
 suggestions for updates on the evolution of 
 $x_{\rm e}$ is an urgent task.  We need to determine that
 the phenomenological parameters $F_{\rm H}$ and $b_{\rm He}$
 are fully understood at the $\lesssim 1$ per cent level before we 
 can be confident that the uncertainties in the details of 
 recombination will have no significant effect on the
 determination of cosmological parameters from {\sl Planck}.


\section{Acknowledgements}
This work made use of the code C{\sc osmo}MC, written by
Antony Lewis and Sarah Bridle.  
We are also grateful to Chris Hirata and Eric Switzer 
for comparing our results with their own calculations.
This research was supported by the Natural Sciences 
and Engineering Research Council of Canada and 
the Canadian Space Agency as well as through a 
University of British Columbia Graduate Fellowship.
A new version of the R{\sc ecfast} code~(version 1.4)
is available from 
http://www.astro.ubc.ca/people/scott/recfast.html.


\bsp

\label{lastpage}
\end{document}